\documentstyle[11pt,paspconf]{article}

\begin{document}

\title{Search for Astronomical Sites in Developing Countries and their
Preventive Protection}
\author{Fran\c{c}ois R. Querci and Monique Querci}
\affil{Observatoire Midi-Pyr\'en\'ees, 14 Av. E. Belin, 31400 Toulouse,
France\break e-mail: querci@ast.obs-mip.fr}

\begin{abstract}
The archives of meteorological satellites permit to find around the
world dry sites well adapted to astronomical observations (in the
visible, IR and millimetric ranges), as a pre-selection of sites.\\
The GSM (Grating Scale Monitor) technique permits to qualify some of
them as astronomical sites for a future setting-up of astronomical
observatories. Such sites are found in new astronomical countries or in
Developing Countries. At the same time, their preventive protection from
light pollution and/or radio interference has to be viewed.\\
In practice, once the pre-selection is made, the governments of these
countries ought to be alerted for example by IAU and/or UN Office for
Outer Space Affairs. The local site testing through GSM should be carried 
out in cooperation with astronomers or scientists of these countries under 
the umbrella of IAU. This should be an approach to help to introduce 
astronomy and astrophysics in Developing Countries.
\end{abstract}


\section{Introduction}
The knowledge of short time-scale variation stars has made some 
progress due to campaigns of non-stop observations with simple manual
telescopes and photometers, conducted simultaneously around the world.
During the last two decades, the automation of telescopes
(remote-controlled or robotic telescopes) opened the way to study the
variability in the entire HR diagram ({\it e.g.} Henry, 1999) and mainly 
at places where simultaneous variations of different characteristic times 
from hours to years are discovered (AGB, RGB stars, etc.). To follow these 
permanent or non-permanent variations during months or years, and to 
understand their origin, networks of robotic telescopes seem to be nowadays 
the most appropriate technology.\\
The first discoveries of the visible counterparts of gamma-ray
bursts and of many new NEO demonstrate that the networks of robotic
telescopes are/will be powerful in many scientific fields ({\it e.g.}
ROBONET, GNAT, TORUS, NORT, etc., as described in Querci and Querci, 1999).\\
Consequently, sites for networks have to be implemented on various
longitudes (continents) and in the two hemispheres. Excellents sites
(Hawaii, North of Chile, South Pole, etc.) or some very good ones
(Canary Islands, South Africa, India, Uzbekistan, etc.) are already at
work. {\bf Are we sure that other excellent sites do not exist elsewhere?} 
In Developing Countries (hereafter DCs) for example?\\
Astronomy and Space Science could be a contribution to the development
of DCs, as already seen, some decades ago, in Canary Islands, Chile, and
so on.

\section{A way to Select Sites in DCs}
A world-wide preliminary map of mean nebulosity (at 0.55 {$\mu$}m) was 
obtained from 12-year meteorological archives and with 250 km square 
meshes (Querci and Querci, 1998a,b).\\

$\bullet$ A first step should be a cross correlation analysis between a
worldwide map of high mountain summits (altitude: 2400-3200 m or more)
and a worldwide map of small-mesh meteorological archives (2 to 5 km) on
cloudiness, humidity, sand winds and light pollution. It could permit to
obtain 20 to 30 new meteorological excellent and/or very good sites
adapted to optical, IR, or millimetric observations.\\

$\bullet$ A second step should be a detailed analysis of the local
atmospheric turbulence for these 20 to 30 pre-selected sites by a 
seeing-monitor or by a grating scale-monitor (G.S.M.) technique 
(Martin et al., 1994). The registration of the parameters L$_{\small 0}$,
the wavefront outer scale, r$_{\small 0}$, the Fried parameter, {$\tau$}, 
the speckle lifetime, and of the isoplanetism angle for each of these 
sites should permit to select finally 8 to 10 sites besides those already 
classified as high quality observing sites.\\

$\bullet$ A third step should be the development of the cooperation and
the ana\-lysis of the local facilities to implement and to maintain
robotic telescopes and their equipment.\\

The two last steps could be a way to introduce Astronomy and Space
Science in DCs through robotic equipment and the analysis of variable
objects supported by hydrodynamical calculations.\\

Preliminary works on large meshes are in progress in many countries.
Moreover, analysis with small meshes are in progress on High Atlas
(Morocco) and on Lebanese border mountains (Syria)(private
communication).

\section{A Protection for the Selected Sites}
In many DCs, Astronomy and Space Science are not developed at all, and
the search for sites is ignored. Consequently, potential sites might be
polluted and lost for science in the future.\\

In a few DCs, collaboration with astronomically-developed countries are
under progress. So, the site prospecting and the preventive site
protection are now taken into account by national scientific
authorities, contributing to the scientific and technical development of
the country.

\section{Conclusion}
The prospecting and the preventive protection of potentially future
astronomical sites are very important for Astronomy and Space Science in
the next century. These sites of which excellent ones are in DCs, have 
been suggested from a world-wide mean annual nebulosity map.\\ 

We take the opportunity of this IAU/UN symposium for asking the
questions: could such a prospecting and preventive protection be promoted\\
\begin{itemize}
\item by each DC individually?\\
\item by some DCs or new astronomical countries grouped together inside 
regional astronomical organizations such as the Arab Union for Astronomy
and Space Science (AUASS)?\\
\item by international astronomical organizations such as European Southern 
Observatory (ESO), Cerro Tololo Inter-American Observatory (CTIO), etc.?
\item by the International Astronomical Union?\\
\item by the UN Office for Outer Space Affairs?
\end{itemize}
 
The encouragement and the help of international organizations would be 
certainly decisive for DCs. At the scientific benefit of themselves as well 
as of the international community.\\


\begin{references}
\reference Henry, G.W.  1999, {\it Techniques for automated
high-precision photometry of Sun-like stars}, P.A.S.P. 111, 845
\reference Martin, P., Tokovinin, A., Agabi, A., Borgnino, J., Ziad, A.
1994, {\it G.S.M.: a Grating Scale Monitor for atmospheric turbulence
measurements. I. The instrument and first results of angle of arrival
measurements}, Astron. Astrophys. Suppl. Ser. 108, 173
\reference Querci, F.R., Querci, M. 1998a, {\it A method for searching
potential observing sites}, in A.S.P. Conf. Series on Preserving the
Astronomical Windows, eds. Syuzo Isobe and Tomohiro Hirayama, vol.139,
p.135 
\reference Querci, F.R., Querci, M. 1998b, {\it The network of oriental
robotic telescopes (NORT)}, at the WWW address: {\it
http://www.saao.ac.za/~wgssa/as2/nort.html}
\reference Querci, F.R., Querci, M. 1999, {\it Robotic telescopes and
networks: new tools for education and science}, in Eight UN/ESA Workshop
on basic Space Science: Scientific Exploration from Space, Mafraq,
Jordan, 13-17 March 1999, to be published in Astrophys. and Space
Science (Kluwer)
\end{references}
\end{document}